# Visible Dielectric Metastructure for Generating Ultranarrow-band Perfect Absorption and Bound States in the Continuum

Md Rumon Miah[†] and Hilmi Volkan Demir*[,†,‡]

[†] Department of Electrical and Electronics Engineering, Department of Physics, UNAM − Institute of Materials Science and Nanotechnology and The National Nanotechnology Research Center, Bilkent University, Ankara 06800, Turkey.

[‡] Luminous! Center of Excellence for Semiconductor Lighting and Displays, School of Electrical and Electronic Engineering, Division of Physics and Applied Physics, School of Physical and Mathematical Sciences, School of Materials Science and Engineering, Nanyang Technological University, Singapore 639798, Singapore.

## Abstract

The ability to generate ultranarrow-band perfect absorption and bound states in the continuum (BICs) in the visible spectrum is highly useful and desired for many advanced optical and nanophotonic applications. However, achieving perfect absorption and BICs with a high quality-factor in the visible region remains challenging due to the material background absorption, radiative losses, and fabrication constraints. Here, we propose and demonstrate a dielectric nanophotonic platform that simultaneously achieves perfect absorption (99.0% at 603 nm) and multiple symmetry-protected BICs in the visible by integrating a dielectric metasurface with a distributed Bragg reflector (DBR). The coupling between the metasurface's localized dielectric resonances and the DBR-assisted photonic mode suppresses radiative leakage and enables ultranarrow-band resonances (full-width at half-maximum ~ 0.65 nm) with strong electromagnetic field (|E|) enhancement (up to 16.4 folds) within the DBR stopband. In addition, the metasurface's asymmetric design enables a magnetic multipole-assisted quasi-BIC resonance with polarization-dependent mode degeneracies, and the periodic structure enables Mie resonance-assisted multiple BICs. Furthermore, the tunability of design parameters provides additional degrees of freedom to control the resonances and their spectral positions. The coexistence of perfect narrowband absorption and BICs within a compact dielectric nanophotonic platform in the visible establishes a promising route toward next-generation low-loss, high-performance meta-optical devices and technologies.

## Introduction

The pursuit of a narrowband, high quality-factor perfect absorption in the visible has attracted significant attention in recent years due to its broad applications in sensors[1,2], solar cells[3,4], photocatalysis[5], photodetectors[6], thermal emitters[7], and lasers[8,9]. One strategy to achieve perfect absorption in extremely thin materials is to block transmission using a reflective surface and vanishing reflectance by introducing interference[10,11,12]. The simplest approach to achieving near-perfect absorption is to stack metal-insulator-metal (MIM) structures to enhance absorption in one metallic layer by minimizing reflections from the other metallic layer[13,14,15]. However, this approach shows relatively broad absorption and works only for a specific polarization state and angle of incidence, which is undesirable for many applications[13]. Another conventional approach is to integrate a thin absorbing layer with a periodic dielectric distributed Bragg reflector (DBR) to create a Tamm plasmonic polariton[16,17,18]. Besides, resonant absorption via critical coupling between radiative and nonradiative resonances in a mirror-spacer-resonator is a widely used approach for achieving near-perfect absorption[19,20]. Although conventional approaches were dominated by plasmonic systems, their inherent Ohmic losses lead to significant thermal dissipation and broad resonance linewidths[1,21,22]. In addition, the low melting points of conventional metals (Au/Ag) are unsuitable for high-temperature applications and intense light illumination[23,24]. To address the inherent limitations of the plasmonic system, using high-refractive-index dielectric materials (Si, $TiO_2$, $Al_2O_3$) has received widespread attention for their low nonradiative losses and ability to generate magnetic multipoles[25,26,27].

Recently, employing optical bound states in the continuum (BICs) in the dielectric metasurfaces has emerged as a powerful tool for realizing ultrahigh-Q resonances[28,29,30,31]. BICs are the discrete states that are completely decoupled from the radiation continuum despite spectrally residing within the continuum[32,33]. Theoretically, BICs possess infinite radiation lifetimes and extremely strong electromagnetic confinement within the structure for the ideal condition, where the structure requires at least one dimension extending to infinity[34,35]. However, in practice, the BICs become quasi-BICs with a very high-Q factor due to the finite dimensions of the structure, material absorption, and external perturbations[36,37,38]. The exceptionally high-Q factor of quasi-BICs has been exploited for many uses, including nonlinear harmonic generation[39,40], photoluminescence enhancement[41], lasing[42,43], sensing and imaging[44,45], vortex generation[46], optical switching[47], and photodetection[48]. Notably, several studies on BIC-driven perfect absorption metasurfaces integrated with metallic reflectors in the near-infrared and mid-infrared regions have been reported[49,50,51].

Despite significant advances in perfect absorption and BICs in the visible, the combined experimental realization of these phenomena via the integration of a dielectric metasurface and a DBR has not yet been explored. In addition, achieving ultranarrow linewidths and perfect absorption simultaneously from a dielectric metasurface in the visible remains highly challenging[25,26]. Furthermore, most reported BIC-based metasurfaces in the visible are designed primarily for reflection or transmission spectra rather than for absorption in the DBR stopband[39,40,41,44].

In this work, we propose and demonstrate a DBR-integrated dielectric nanophotonic platform that supports ultranarrow-band perfect absorption (99.0% at 603 nm) along with BICs in the visible spectrum. Our design achieves highly confined optical modes with strongly suppressed radiative leakage by interacting the localized dielectric resonances from the low-loss metasurface with the strong vertical-mode confinement provided by the DBR. The asymmetric meta-atom and its symmetric counterpart, with their dependence on the polarization direction of the incident source, enable control over absorption resonances. The proposed dielectric architecture minimizes Ohmic dissipation, enabling a high-Q (~928) and an ultranarrow absorption resonance (FWHM ~ 0.65 nm) while preserving strong electromagnetic field (|E|) enhancement (up to 16.4 folds).

Our periodic design leverages multiple symmetry-protected BICs, and the asymmetric meta-atom enables magnetic-multipole-assisted quasi-BIC resonance with polarization-dependent mode degeneracies within the DBR stopband. Besides, the tuning of the meta-atom's asymmetry difference and the spacer thickness facilitates the control over the selectivity of resonances and the spectral position. In addition, we demonstrate the BICs in the reflection spectrum, free of absorption, using the identical metasurface design by replacing the total-reflective background with a transmissive one to validate the DBR-integrated hybrid design. The coexistence of narrowband perfect absorption and multiple BICs within a compact visible-frequency dielectric platform holds great promise for the next-generation implementation of visible meta-optics and nanophotonic technologies.

## Results and discussion

The integration of the DBR with the metasurface enables cavity modes between them, which, coupled with the metasurface's excited dielectric resonances, provide absorption resonances within the DBR stopband region with unity efficiency and stronger field confinement[20,52,53,54]. Besides, the DBR can completely suppress radiation into the substrate, thereby effectively increasing the Q-factor of excited modes from the metasurface[52]. In addition, the cavity modes themselves generate high-Q resonances that accumulate optical energy within the device[17,20]. The asymmetric T-shaped hole is designed to introduce symmetry breaking and a difference in the displacement current density across the two sides of the meta-atom, thereby exciting magnetic resonances that give rise to the quasi-BIC.

Our proposed design concept is depicted in Figure 1a. Here, T3 is the thickness of the critical spacing layer (480 nm-thick $SiO_2$) inserted between the metasurface and the DBR. The purpose of this spacing dielectric layer is to create the Fabry–Pérot (FP)-like cavity modes between the metasurface and the top layer of DBR. In addition, it prevents the metasurface from interacting with the DBR's near field. On the other hand, the dielectric coating layer (T1) made of polymethyl methacrylate (PMMA, n = 1.49) provides a uniform refractive index environment for the metasurface by matching its optical constants to those of the underlying $SiO_2$ layer. The red meta-atom in Figure 1a shows the top view of the unit cell with the regular arm width of 105 nm, while the thicker arm width is 165 nm. The asymmetric arms start from the halfway point of the meta-atom. All arms become 105 nm wide for the symmetric square hole shape with a side length (q) of 200 nm. For both designs, the metasurface thickness (T2) is fixed at 240 nm, and the entire device is fabricated on a fused silica substrate. The total thickness of the DBR (T4) is 1,376 nm. We employed finite-difference time-domain (FDTD) calculations (Lumerical) for all our numerical simulations of this proposed architecture.

To fabricate the proposed structure, we deposited $TiO_2$ using ultra-high-vacuum RF sputtering, while $SiO_2$ was deposited by temperature-controlled plasma-enhanced chemical vapor deposition (PECVD). The metasurface was patterned using electron beam lithography (EBL), followed by dry etching in inductively coupled plasma (ICP). The top-view scanning electron microscopy (SEM) images of the fabricated asymmetric T-shaped hole and corresponding square hole metasurface are presented in Figures 1b and 1c, for which details of the fabrication procedure are described in the methods section. We used a 4f lens-based back focal plane (BFP) imaging technique for all experimental measurements (details given in the methods section). Figure 1d shows the numerical reflectance of the DBR, where each layer is calculated by dividing the quarter wavelength by the corresponding material's refractive index, where the central wavelength is set to 620 nm. The maximum numerical reflectance from the DBR reaches at 99.8% after eight pairs of $SiO_2/TiO_2$ layers with a spectral range of 168 nm (90% reflectivity from 532 to 700 nm). The fabricated DBR with 8 periods exhibits the same maximum reflectance of 99.8% over a spectral range of 177 nm (520 to 697 nm). The vertical cross-sectional SEM image of the device is presented in Supplementary Information S1.

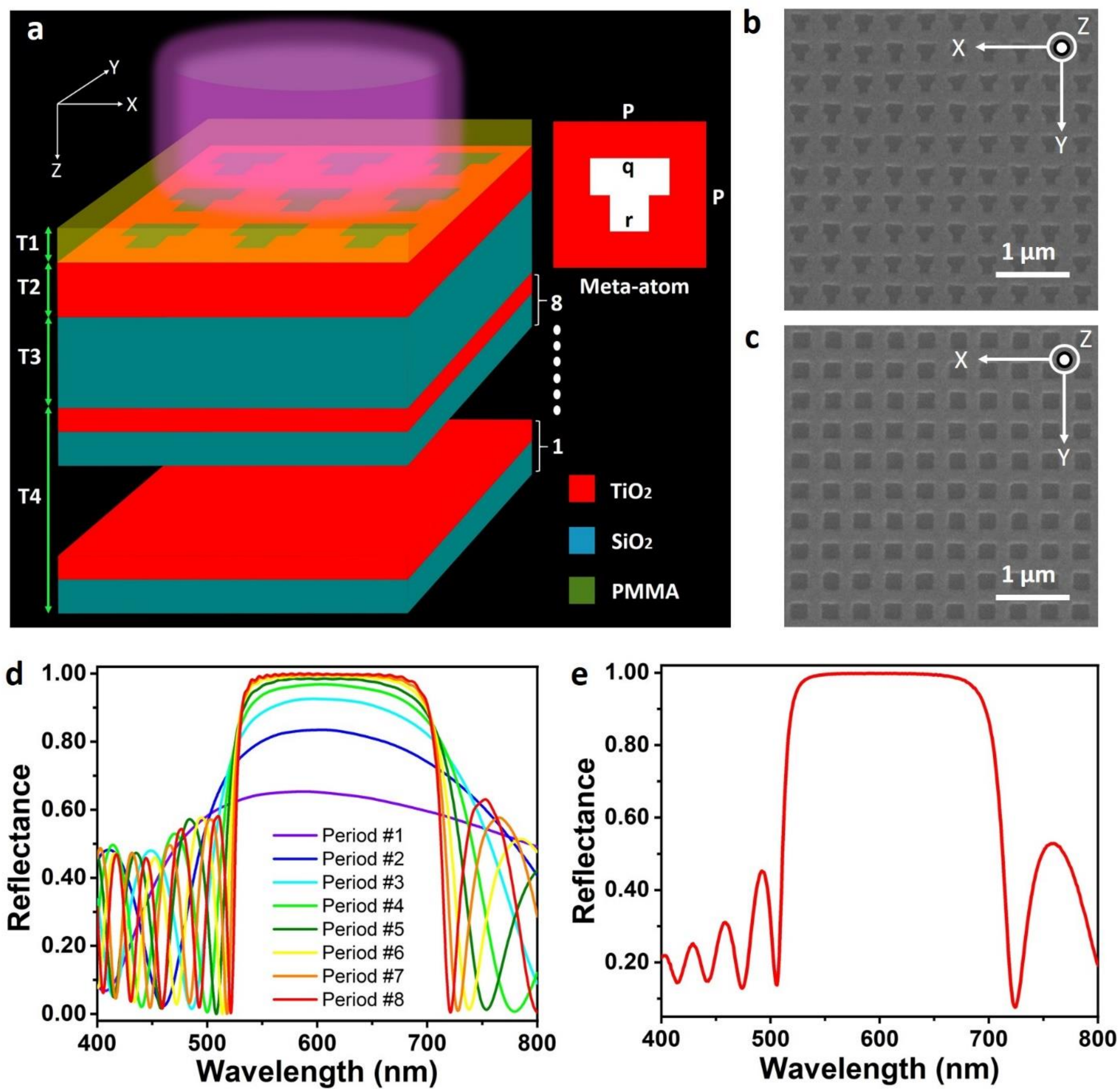

**Figure 1. Structural design, SEM images, and numerical and experimental reflectance from DBR. a.** Schematic illustration represents the proposed device. T1, T2, and T3 denote the thickness of the coated PMMA film, metasurface layer, and inserted fused silica layer (480 nm) between the metasurface and DBR, respectively. T4 shows the DBR structure with 8 pairs of alternating $SiO_2$ (104 nm) and $TiO_2$ (68 nm) layers. Meta-atom represents the unit cell of the metasurface, where period, p = 410 nm, q = 200 nm, and r = 80 nm. The z-axis denotes the propagation direction of the applied plane wave source. **b, c,** Top-view scanning electron microscope (SEM) images of the fabricated asymmetric T-shaped hole and corresponding square hole metasurfaces, respectively. **d,** Numerical reflectance spectra from the DBR, with increasing number of repeating periods of $SiO_2$ and $TiO_2$ thin films from 1 to 8. **e,** Experimental reflectance spectrum from the fabricated DBR with 8 periods.

Predominantly, the resonances that emerge in the metasurface's scattering spectrum are peaks when integrated with a transmissive background, such as glass. However, the resonances can be sharp dips when the background changes to a highly reflective structure, such as a DBR or a metallic mirror. Fabricating a metallic mirror is straightforward; however, the strong Ohmic losses in the visible region are an obstacle to achieving high reflectivity. In contrast, a multi-layered dielectric DBR offers almost unity reflectance over a wider spectral range, and the spectral position of the reflectance is tunable by modulating the layer thicknesses and the optical constants of the dielectric materials. Figure 2 demonstrates the numerical and normalized experimental angle-resolved reflectance and numerical reflectance spectra for the normally applied source after the integration of the dielectric metasurface with the DBR, where all resonances appear as the sharp dips in the reflectance spectra.

Figure 2a shows several highly narrowband absorption resonances, with two resonances at 603 and 619 nm having Q-factors of 928 (FWHM ~ 0.65 nm) and 768 (FWHM ~ 0.81 nm), respectively, and achieving high absorption levels of 99.0% and 93.3%, respectively, and are very stable with respect to the angle of incidence of the source. In contrast, two additional resonances at 605 and 622 nm are very weak as a function of the source angle. Actually, the resonances at 603 and 605 are the same electric dipole (ED) resonances, degenerate into two modes under x- and y-polarization due to the metasurface's symmetry breaking. Similarly, the magnetic quadrupole (MQ) mode resonance splits into two at 619 and 622 nm. More importantly, there is a discernible BIC point at 609 nm (inset of Figure 2a) due to the in-plane symmetry against the normally applied periodic source, which appears as a quasi-BIC resonance dip while the angle of incidence is tuned.

The normalized experimental angle-resolved reflectance (Figure 2b) measured by BFP imaging supports the numerical results, in which weak resonances are undetected, and some resonances are spectrally shifted due to limitations of the characterization system and fabrication imperfections. The numerical reflectance spectra in Figure 2c demonstrate the wavy reflectance above and below the off-resonance reflectance, because of the DBR-integrated modified photonic environment. The integration of DBR with the dielectric metasurface creates Tamm-like localized surface states, resulting in oscillatory scattering spectra within the DBR stopband. The MQ mode resonance at 619 is significantly reduced when the polarization angle of the source is rotated by 90°, as shown in Figures 2d and 2f. This occurs because the rotation of the applied electric field from the asymmetric to the symmetric axis results in a minimum difference in the displacement current density into the meta-atoms, thereby suppressing the magnetic multipoles. Furthermore, there are two symmetry-protected BICs at 622 and 568 nm, which are obvious from Figure 2d, and the normalized experimental angle-resolved reflectance of Figure 2e supports this observation. The BIC at 622 nm arises from the decoupling of the MQ from the applied source, and the BIC at 568 nm arises due to the in-plane symmetry of the higher-order mode against the normally applied periodic source.

The reflectance spectra from the symmetric square hole metasurface after the integration with the DBR demonstrate only one sharp resonance dip at 601 nm for both polarizations of a normally applied source (Figure 2i). Due to the symmetric design, no resonance splitting happens for the ED mode at 601 nm. Besides, the MQ mode resonance at 619 nm is also undetected as the structure turns to a symmetric design. At this point, we can conclude that the MQ resonance creates the symmetry-protected BIC for the square hole-shaped design, which we discuss further in the next section. In addition, the angle-resolved reflectance demonstrates the angle-tuned BICs at 606 nm (Figure 2g) and at 610 and 614nm (Figure 2j). The normalized experimental angle-resolved reflectance of Figures 2h and 2k supports their corresponding numerical results. Another, weaker resonance, which is stable against asymmetry and the angle of the applied source, is observed at 636 nm (Figures 2c, 2f, and 2i) due to the FP cavity mode. Furthermore, all experimental angle-resolved reflectance spectra exhibit wavy noise due to the FP cavity effect between the DBR and the metasurface. The significant advantage of this novel design is the presence of ultra-narrowband absorption resonances with several BICs, which can be controlled by the polarization directions and design geometry.

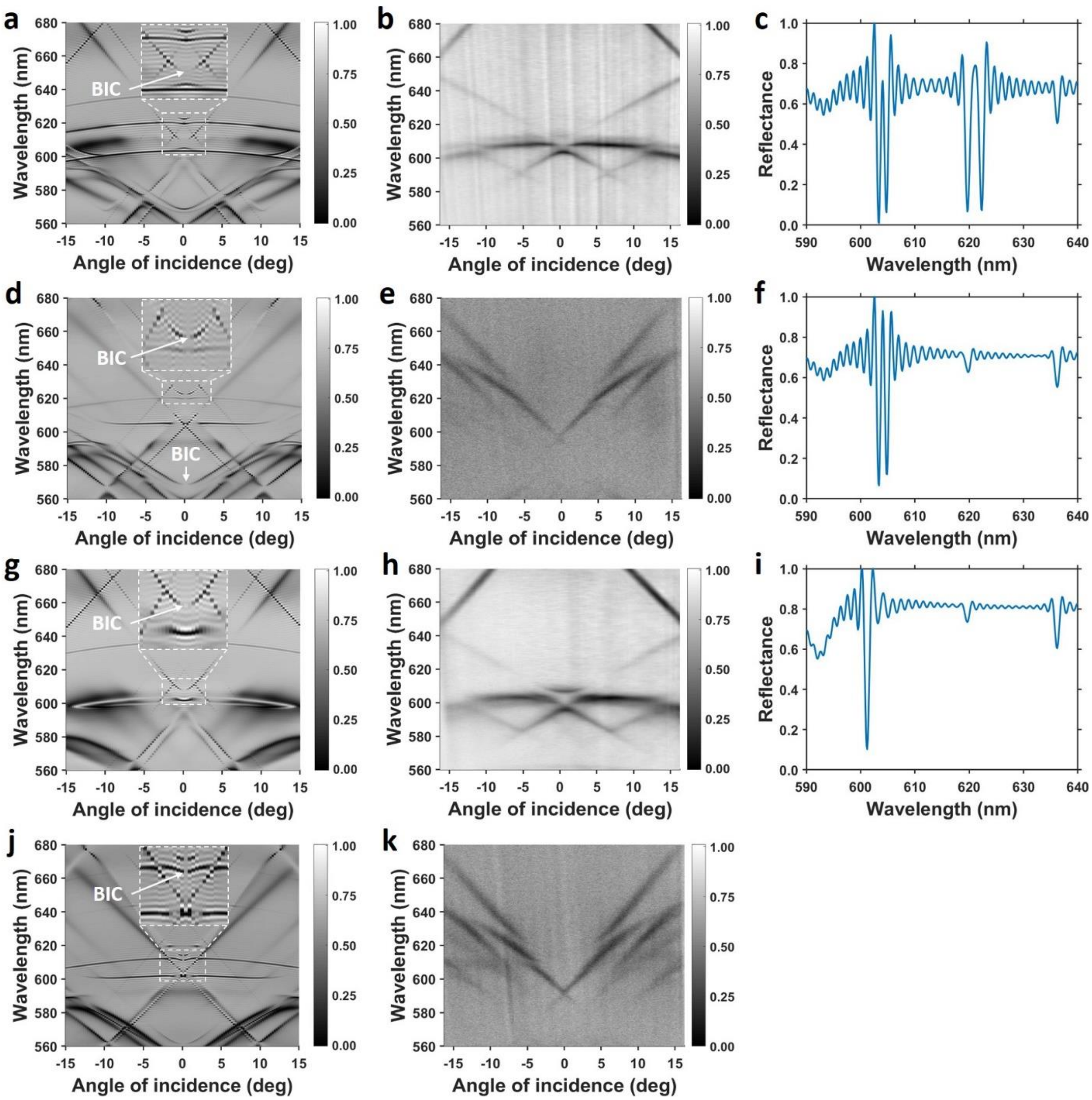

**Figure 2. Angle-resolved reflectance and reflectance for the normally applied source from the DBR-integrated metasurface. a, b,** Numerical and experimental angle-resolved reflectance from the asymmetric T-shaped hole metasurface for the polarization direction of the applied source along the x-axis. **c,** Numerical reflectance for the normally applied source from Figure 2a, representing very high-Q absorption resonances. **d, e,** Numerical and experimental angle-resolved reflectance from the asymmetric T-shaped hole metasurface for the polarization direction along the y-axis. **f,** Numerical reflectance for the normally applied source from Figure 2d. **g, h,** Numerical and experimental angle-resolved spectra from the corresponding square hole metasurface, while the polarization direction of the applied source is along the x-axis. **i,** Numerical reflectance for the normally applied source from Figure g. **j, k,** Numerical and experimental angle-resolved reflectance spectra from the square hole metasurface for the polarization along the y-axis.

To confirm the existence of quasi-BIC resonances in Figure 2a, and the existence of BICs in Figures 2g and 2h, we performed the tuning of the asymmetry difference between the thicker arm and the regular arm of the designed T-shaped meta-atom. Three quasi-BIC resonances appear at 568, 619, and 622 nm in Figure 2a for the asymmetric T-shaped design, which are present in Figure 3a for the asymmetry difference of 60 nm. These three quasi-BICs gradually decrease their intensity and spectral width as the asymmetry difference decreases, and eventually, BIC points are pushed to 561, 610, and 614 nm (Figure 3a) for the case of the symmetric design (square shape). When the polarization direction rotates to the y-axis, the quasi-BIC resonances completely disappear for any asymmetric difference (Figure 3b). Besides, the ED resonance at 603 nm shifts slightly and splits into two resonances after having a large asymmetric difference. Unlike ED, MQ at 619 and 622 nm maintains the splitting over the entire range of asymmetry differences and shifts the spectral position more rapidly due to its strong dependence on the asymmetric shape. Besides, the resonance at 636 nm in Figure 3 is independent of the asymmetry difference, as it arises from the cavity mode.

In addition to the asymmetry difference, we also tuned the spacer thickness from 0 to 600 nm to validate the designed thickness of 480 nm. At 480 nm, the reflectance shows high-Q absorption resonances for both ED and MQ, with resonance splitting (Figure 3c), and the MQ resonance diminishes significantly as the polarization direction is rotated by 90° (Figure 3d). The MQ absorption resonances blue-shifted slightly as the spacer thickness decreased from 400 nm, and the splitting disappeared, and the intensity reduced significantly after 120 nm (Figure 3c). Meanwhile, the ED resonance is identical for both polarizations and blue-shifts gradually as the spacer thickness decreases. Additionally, all resonances in Figures 3c and 3d exhibit a periodic intensity discontinuity with respect to the spacer thickness, validating the FP mode. In addition, the reflectance is polarization-invariant in the absence of the spacing layer, with several additional resonances due to the DBR's near-field effect. The numerical reflectance for the device without a spacing layer is presented in Supplementary Information S2.

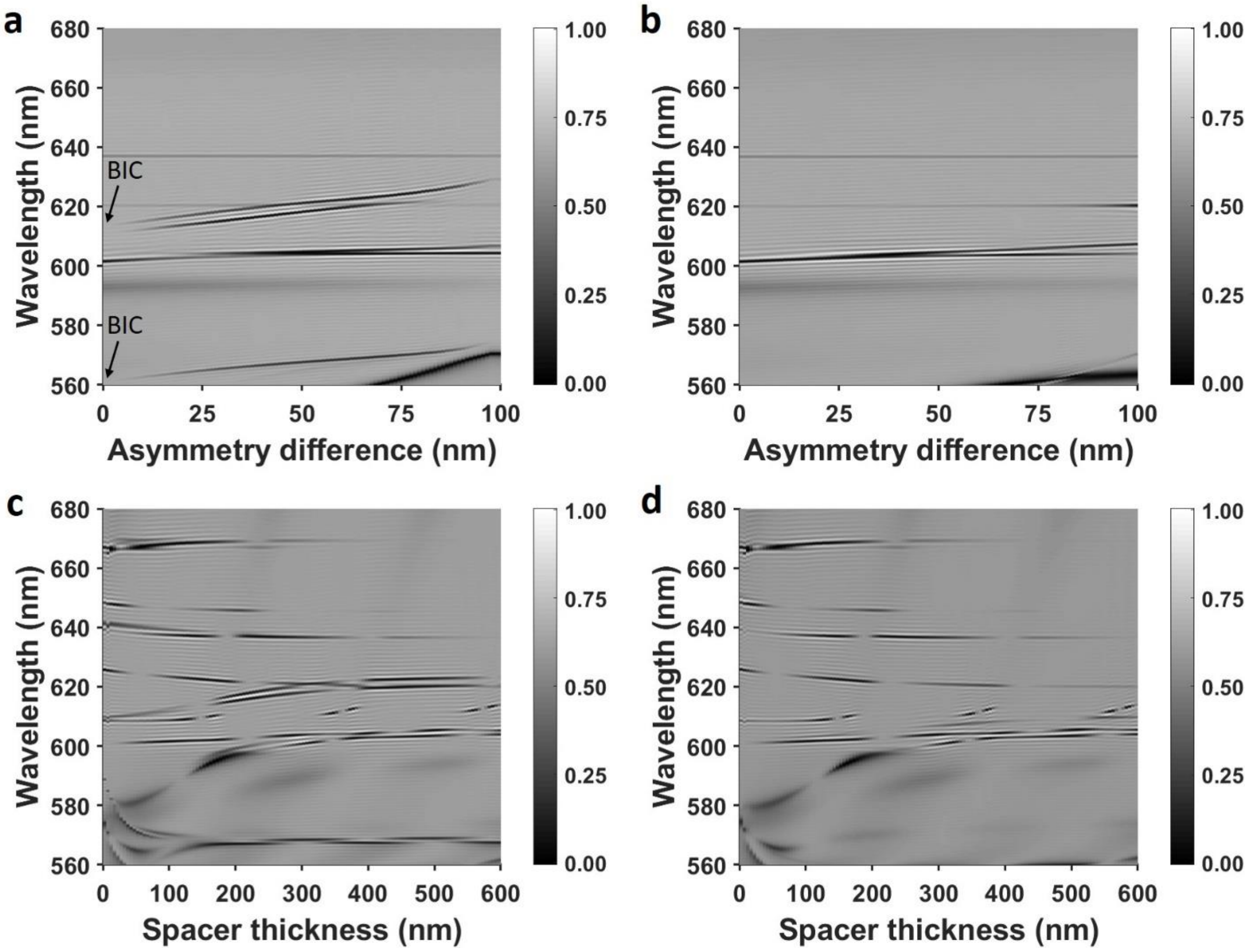

**Figure 3. Numerical reflectance for the tuning of the asymmetry difference and spacer thickness. a, b,** Reflectance for the asymmetric difference tuning between the thicker arm and the regular arm of the meta-atom while the polarization of the applied source is along the x-axis (a), and the y-axis (b). The meta-atom turns into a square-shaped hole at the minimum asymmetry difference, where three BIC points (614, 610, and 561 nm) appear due to the symmetry protection (a). **c, d,** Reflectance for the tuning of spacer thickness between the DBR and metasurface while the polarization of the applied source is along the x-axis (c), and the y-axis (d). Reflectance for spacer tuning demonstrates control of the resonance at 480 nm.

We investigated the electric field distribution maps from the vertical cross-section for the asymmetric T-shaped structure to support our claim of the generation of resonances and their coupling with the cavity mode. Figure 4 shows the field distribution maps and corresponding field enhancements for wavelengths of 603, 619, 636, and 630 nm, respectively, from Figure 2c. The ED at 603 nm couples to the cavity mode (Figure 4a); hence, it shows strong spectral invariance with respect to the asymmetry difference (Figure 3) and the propagation angle of the applied source (Figure 2). In contrast, the quasi-BIC resonance at 619 nm shows complete decoupling from the cavity mode (Figure 4b), with a field enhancement factor of around 16.4 in the metasurface region, validating that the dielectric resonance (MQ) arises from the metasurface only. The absorption resonance at 636 nm demonstrates the FP cavity mode between the metasurface and the bottom layer of the DBR (Figure 4c), which supports the spectral invariance of the resonance against asymmetry differences (Figure 3) and incident-light angles (Figure 2). In addition to the electric field from resonance dips, we also investigated the electric field map in the continuum region at 630 nm (Figure 4d). The map demonstrates the complete reflection of the applied field from the DBR, and no interaction with the metasurface. The electric field distribution maps from the horizontal and vertical cross-sections for all designs and polarization directions, along with associated field enhancements, are presented in Supplementary Information S3. The electric field vector distribution maps and multi-mode decomposition analysis are presented in Supplementary Information S4.

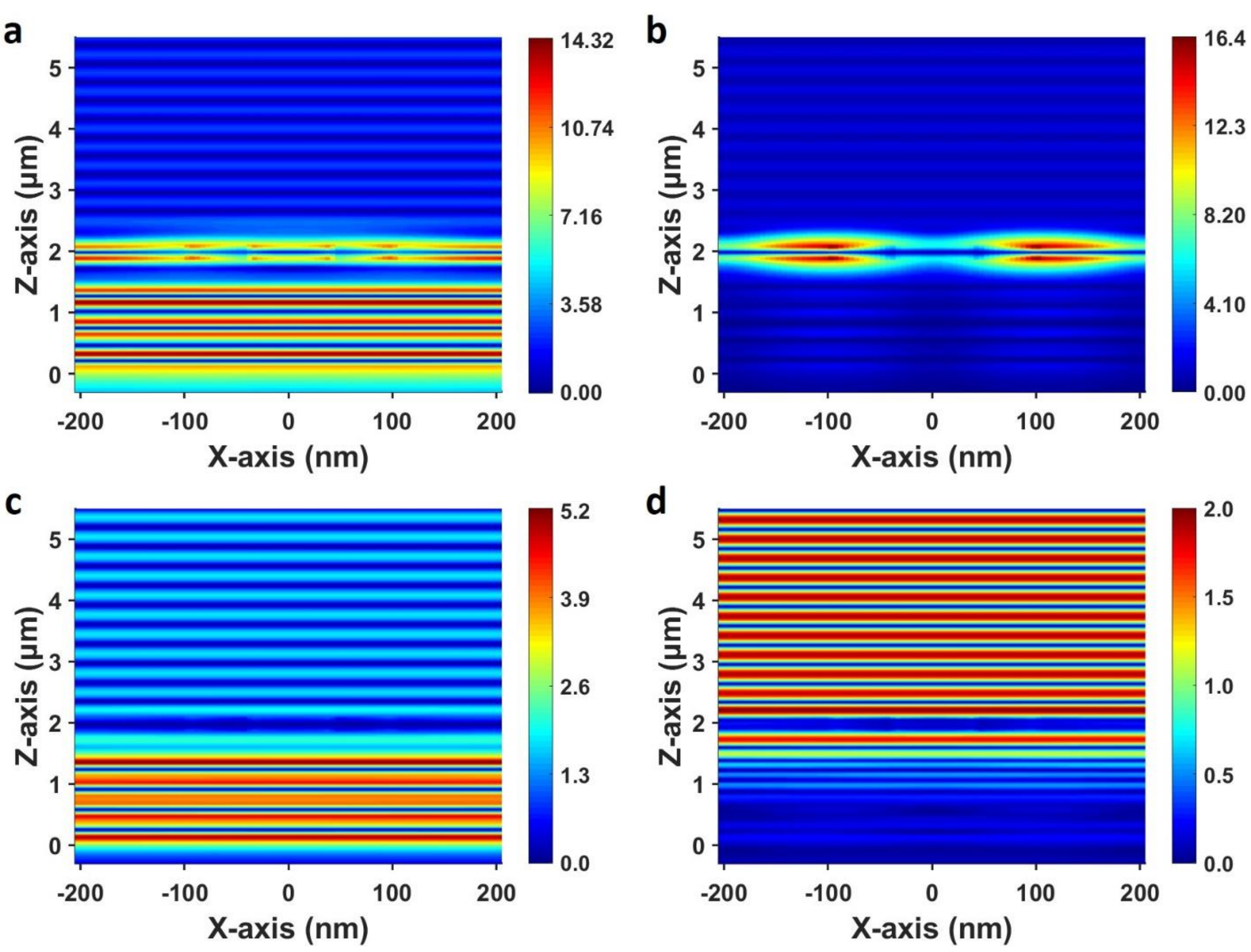

**Figure 4.** Electric field distribution maps from the vertical cross-section for the asymmetric T-shape hole, where the polarization direction of the applied source is along the x-axis. The DBR, dielectric spacer, and metasurface layers start from 0, 1.376, and 1.851 μm, respectively. **a,** Map from the absorption resonance dip at 603 nm, representing the coupling between the ED and cavity mode. **b,** Map from the quasi-BIC absorption dip at 619 nm, representing the heavy field confinement and weakly coupled with the continuum. **c, d,** Maps from the resonance dip at 636 nm (c) showing the FP cavity mode, and continuum mode at 630 nm (d), respectively.

We investigated the metasurface's reflection with and without DBR integration to understand the effect of DBR on the metasurface. We simulated and fabricated meta-atoms with the exact dimensions of the DBR and spacer integrated for both the asymmetric T-shaped and the symmetric square-hole metasurfaces on fused silica substrates, which were both covered by an identical thick PMMA layer. The numerical and experimental angle-resolved reflectance spectra without a DBR and spacer (Figure 5) show fewer, broader resonances and fewer BICs. The transmissive background (fused silica) provides a wide passband with narrow resonance stopbands, which is the opposite of the DBR-integrated reflection spectrum. The asymmetric T-shaped hole exhibits Fano-type resonances at 619 and 622 nm for the applied source along the x-axis (Figure 5c), which become symmetry-protected BICs when the applied source is rotated along the y-axis (Figure 5f), and the structure turns to a symmetric square hole (Figure 5i). In addition, the angle-dependent quasi-BIC resonances converge to the BICs at 622 nm (Figure 5d), 610 nm (Figure 5J), and 614 nm (Figure 5J), respectively. Similarly, two BICs at 609 nm (Figure 5a) and 606 nm (Figure 5g) maintain spectral positions similar to those of the DBR-integrated design. Therefore, we can conclude that the origin of these BICs is independent of DBR.

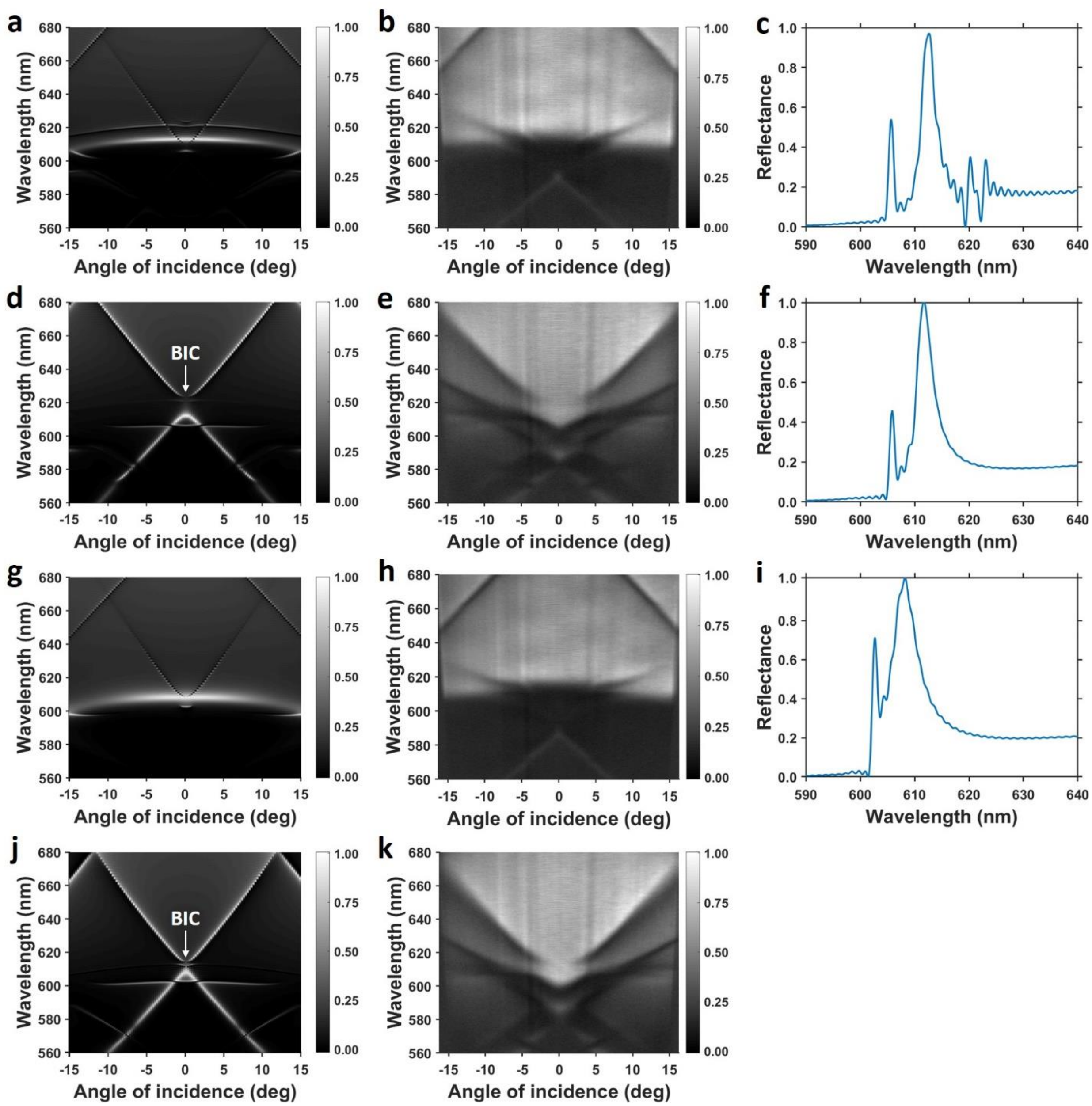

**Figure 5. Angle-resolved reflectance and reflectance for the normally applied source from the metasurface (without DBR). a, b,** Numerical and experimental angle-resolved reflectance from the asymmetric T-shaped hole metasurface for the polarization direction of the applied source along the x-axis. **c,** Numerical reflectance for the normally applied source from Figure a, representing two reflective resonances and two Fano-type resonances. **d, e,** Numerical and experimental angle-resolved reflectance from the asymmetric T-shaped hole metasurface for the polarization direction along the y-axis. **f,** Numerical reflectance for the normally applied source from Figure d. **g, h,** Numerical and experimental angle-resolved spectra from the corresponding square hole metasurface, while the polarization direction of the applied source is along the x-axis. **i,** Numerical reflectance for the normally applied source from Figure g. **j, k,** Numerical and experimental angle-resolved reflectance spectra from the square hole metasurface for the polarization along the y-axis.

## Conclusion

In summary, we proposed and demonstrated numerically and experimentally a dielectric nanophotonic platform consisting of a metasurface interacted with a distributed Bragg reflector, which enables simultaneous narrowband perfect absorption and bound states in the continuum in the visible. The proposed architecture achieved near-unity absorption with a maximum FWHM of ~ 0.65 nm while maintaining strong field enhancement (up to 16.4 folds) through coupling between the dielectric resonance modes of the metasurface and the DBR-assisted photonic mode. In addition, the proposed device demonstrated multiple symmetry-protected BICs via dielectric resonances arising from its periodic Bloch structure, and it excited a MQ-assisted quasi-BIC in the broken-symmetry structure. We also developed the understanding of the design parameter tuning and polarization dependence, which provide control over the absorption resonance and spectral selectivity. The numerical and experimental results with BICs from the metasurface without the reflective background support the hybrid design strategies. Furthermore, we analyzed the mechanism underlying the resonances to gain deeper insight, presenting electric-field distribution maps and the interaction with the DBR. Finally, the hybrid, compact, DBR-integrated nanophotonic platform could be a compelling route to advanced meta-optics and next-generation photonic technologies.

## Methods

**Numerical calculations.** The numerical reflectance, angle-resolved reflectance, and field maps were calculated employing the finite-difference time-domain (FDTD) based commercially available software Lumerical. Periodic and perfectly matched layer (PML) boundary conditions were applied in the horizontal and vertical directions to match the periodic structure and absorb reflected and transmitted light. The Bloch/periodic wave source was applied as a normally applied source, while the broadband fixed angle source technique (BFAST) source type was employed for the angle illumination. Additionally, the Frequency-domain field and power monitors were inserted through the horizontal and vertical midpoints of the metasurface to obtain the field maps.

**Fabrication.** The standard cleaned fused silica wafers using Piranha process were placed in the plasma-enhanced chemical vapor deposition (PECVD) and RF-sputter chambers to deposit $SiO_2$ and $TiO_2$ thin films. The PMMA (e-beam resist) was spin-coated on top of the final $TiO_2$ layer, followed by a very thin layer (7 nm) of spin-coated conducting polymer espacer 300Z (SHOWA DENKO) to avoid charge accumulation during the electron beam lithography (EBL) process. After patterning the target design using EBL, development was carried out in MIBK : IPA (1:3), IPA, and DI water. Then, a chromium (Cr) layer (50 nm) was deposited as a hard mask employing the e-beam evaporator (MIDAS e-beam evaporator). After that, the lift-off process was performed by rinsing in acetone in an ultrasonicator bath to remove the unwanted PMMA and Cr. Next, the hard-mask pattern was transferred to the target $TiO_2$ layer via inductively coupled plasma (ICP) etching (STS Multiplex ICP) using $CHF_3$ (20 sccm), $SF_6$ (20 sccm), and $O_2$ (4 sccm) gas flows. Finally, the chrome etchant is used to wet etch the remaining Cr hard mask. A PMMA layer was deposited on the final fabricated structure by spin-coating for optical characterization.

**Optical characterization.** A back focal plane (BFP) optical characterization setup was used to perform angle-resolved reflectance measurements. The BFP setup includes a 4f lens assembly, a halogen lamp connected via an optical fiber, a collimation lens, beam splitters, mirrors, a Mitutoyo Plan Apo 10× objective (NA 0.28), a polarizer, a spatial filter, a slit spectrometer, and a CCD camera.

## Author information

### Corresponding Author

**Hilmi Volkan Demir** – LUMINOUS! Centre of Excellence for Semiconductor Lighting and Displays, School of Electrical and Electronic Engineering, The Photonics Institute (TPI), Nanyang Technological University, 639798, Singapore; School of Physical and Mathematical Sciences, Division of Physics and Applied Physics, Nanyang Technological University, 639798, Singapore; Department of Electrical and Electronics Engineering and Department of Physics, UNAM - Institute of Materials Science and Nanotechnology, Bilkent University, Ankara 06800, Turkey; orcid.org/0000-0003- 1793-112X; Email: volkan@bilkent.edu.tr, hvdemir@ntu.edu.tr

### Author

**Md Rumon Miah -** Department of Electrical and Electronics Engineering and Department of Physics, UNAM - Institute of Materials Science and Nanotechnology, Bilkent University, Ankara 06800, Turkey; orcid.org/0000-0001-5247-3967

### Notes

The authors declare no competing financial interest.

## Acknowledgement

The authors gratefully acknowledge the financial support from TUBITAK 20AG001 and 121C266. H. V. D. also acknowledges the support from TUBA.

## References

1. Tittl, A. *et al.* Palladium-based plasmonic perfect absorber in the visible wavelength range and its application to hydrogen sensing. *Nano Lett.* **11**, 4366–4369 (2011).
2. Bhattarai, K. *et al.* A Large-Area, Mushroom-Capped Plasmonic Perfect Absorber: Refractive Index Sensing and Fabry-Perot Cavity Mechanism. *Adv. Opt. Mater.* **3**, 1779–1786 (2015).
3. Chaudhry, F. A. *et al.* Light absorption enhancement in thin film GaAs solar cells using dielectric nanoparticles. *Sci. Rep.* **12**, 1–9 (2022).
4. Zhang, C. *et al.* Absorption enhancement in thin-film organic solar cells through electric and magnetic resonances in optical metamaterial. *Opt. Mater. Express* **5**, 1954 (2015).
5. Xiao, Q. *et al.* Hot-Carrier Organic Synthesis via the Near-Perfect Absorption of Light. *ACS Catal.* **8**, 10331–10339 (2018).
6. Sobhani, A. *et al.* Narrowband photodetection in the near-infrared with a plasmon-induced hot electron device. *Nat. Commun.* **4**, 1–6 (2013).
7. He, M. *et al.* Deterministic inverse design of Tamm plasmon thermal emitters with multi-resonant control. *Nat. Mater.* **20**, 1663–1669 (2021).
8. Symonds, C. *et al.* Hybrid metal/semiconductor lasers based on confined Tamm plasmons. *Phys. Simul. Optoelectron. Devices XXII* **8980**, 89800J (2014).
9. Xu, W.-H. *et al.* Tamm Plasmon-Polariton Ultraviolet Lasers. *Adv. Photonics Res.* **3**, 1–6 (2022).
10. Cui, Y. *et al.* Plasmonic and metamaterial structures as electromagnetic absorbers. *Laser Photonics Rev.* **8**, 495–520 (2014).
11. Hajian, H., Ghobadi, A., Butun, B. & Ozbay, E. Active metamaterial nearly perfect light absorbers: a review [Invited]. *J. Opt. Soc. Am. B* **36**, F131 (2019).
12. Chen, H.-T. Interference theory of metamaterial perfect absorbers. *Opt. Express* **20**, 7165 (2012).

13. Kocer, H., Butun, S., Li, Z. & Aydin, K. Reduced near-infrared absorption using ultra-thin lossy metals in Fabry-Perot cavities. *Sci. Rep.* **5**, 1–6 (2015).

14. Serhatlioglu, M., Ayas, S., Biyikli, N., Dana, A. & Solmaz, M. E. Perfectly absorbing ultra thin interference coatings for hydrogen sensing. *Opt. Lett.* **41**, 1724 (2016).

15. Aalizadeh, M., Khavasi, A., Butun, B. & Ozbay, E. Large-Area, Cost-Effective, Ultra-Broadband Perfect Absorber Utilizing Manganese in Metal-Insulator-Metal Structure. *Sci. Rep.* **8**, 1–13 (2018).

16. Lu, H., Gan, X., Jia, B., Mao, D. & Zhao, J. Tunable high-efficiency light absorption of monolayer graphene via Tamm plasmon polaritons. *Opt. Lett.* **41**, 4743 (2016).

17. Lundt, N. *et al.* Room-temperature Tamm-plasmon exciton-polaritons with a WSe2 monolayer. *Nat. Commun.* **7**, (2016).

18. Song, D., Wu, B., Liu, Y., Wu, X. & Yu, K. A polarization-dependent perfect absorber with high Q-factors enabled by Tamm phonon polaritons in hyperbolic materials. *Phys. Chem. Chem. Phys.* **25**, 25803–25809 (2023).

19. Aydin, K., Ferry, V. E., Briggs, R. M. & Atwater, H. A. Broadband polarization-independent resonant light absorption using ultrathin plasmonic super absorbers. *Nat. Commun.* **2**, 1–7 (2011).

20. Huang, S. H. *et al.* Microcavity-assisted multi-resonant metasurfaces enabling versatile wavefront engineering. *Nat. Commun.* **15**, 1–11 (2024).

21. Akselrod, G. M. *et al.* Large-Area Metasurface Perfect Absorbers from Visible to Near-Infrared. *Adv. Mater.* **27**, 8028–8034 (2015).

22. Hedayati, M. K., Zillohu, A. U., Strunskus, T., Faupel, F. & Elbahri, M. Plasmonic tunable metamaterial absorber as ultraviolet protection film. *Appl. Phys. Lett.* **104**, (2014).

23. Pamato, M. G., Wood, I. G., Dobson, D. P., Hunt, S. A. & Vočadlo, L. The thermal expansion of gold: point defect concentrations and pre-melting in a face-centred cubic metal. *J. Appl. Crystallogr.* **51**, 470–480 (2018).

24. Yeshchenko, O. A. *et al.* Size and Temperature Effects on the Surface Plasmon Resonance

in Silver Nanoparticles. *Plasmonics* **7**, 685–694 (2012).

25. Yang, C. Y. *et al.* Nonradiating Silicon Nanoantenna Metasurfaces as Narrowband Absorbers. *ACS Photonics* **5**, 2596–2601 (2018).

26. Buhara, E., Ghobadi, A. & Ozbay, E. An All-Dielectric Metasurface Coupled with Two-Dimensional Semiconductors for Thermally Tunable Ultra-narrowband Light Absorption. *Plasmonics* **16**, 687–694 (2021).

27. Yang, F. *et al.* Ultraviolet narrowband all-dielectric metasurface absorber with an ultra-thin absorption layer. *Opt. Lett.* **50**, 2049 (2025).

28. Koshelev, K., Bogdanov, A. & Kivshar, Y. Engineering with Bound States in the Continuum. *Opt. Photonics News* **31**, 38 (2020).

29. Koshelev, K. L., Sadrieva, Z. F., Shcherbakov, A. A., Kivshar, Y. S. & Bogdanov, A. A. Bound states in the continuum in photonic structures. *Physics-Uspekhi* **66**, 494–517 (2023).

30. Joseph, S., Pandey, S., Sarkar, S. & Joseph, J. *Bound states in the continuum in resonant nanostructures: An overview of engineered materials for tailored applications*. *Nanophotonics*. **10**, 4175-4207 (2021).

31. Koshelev, K., Bogdanov, A. & Kivshar, Y. Meta-optics and bound states in the continuum. *Sci. Bull.* **64**, 836–842 (2019).

32. Hsu, C. W., Zhen, B., Stone, A. D., Joannopoulos, J. D. & Soljacic, M. Bound states in the continuum. *Nat. Rev. Mater.* **1**, (2016).

33. Hsu, C. W. *et al.* Observation of trapped light within the radiation continuum. *Nature* **499**, 188–191 (2013).

34. Azzam, S. I. & Kildishev, A. V. Photonic Bound States in the Continuum: From Basics to Applications. *Adv. Opt. Mater.* **9**, 16–24 (2021).

35. Azzam, S. I. *et al.* Single and Multi-Mode Directional Lasing from Arrays of Dielectric Nanoresonators. *Laser Photonics Rev.* **15**, 1–8 (2021).

36. Liu, Z. *et al.* High- Q Quasibound States in the Continuum for Nonlinear Metasurfaces.

*Phys. Rev. Lett.* **123**, 1–6 (2019).

37. Yuan, L. & Lu, Y. Y. Bound states in the continuum on periodic structures surrounded by strong resonances. *Phys. Rev. A* **97**, 4490–4493 (2018).

38. Sadrieva, Z. F. *et al.* Transition from Optical Bound States in the Continuum to Leaky Resonances: Role of Substrate and Roughness. *ACS Photonics* **4**, 723–727 (2017).

39. Bernhardt, N. *et al.* Quasi-BIC Resonant Enhancement of Second-Harmonic Generation in WS2Monolayers. *Nano Lett.* **20**, 5309–5314 (2020).

40. Anthur, A. P. *et al.* Continuous Wave Second Harmonic Generation Enabled by Quasi-Bound-States in the Continuum on Gallium Phosphide Metasurfaces. *Nano Lett.* **20**, 8745–8751 (2020).

41. Lee, J. *et al.* Bound-States-in-the-Continuum-Induced Directional Photoluminescence with Polarization Singularity in WS2 Monolayers. *Nano Lett.* **25**, 861–867 (2025).

42. Wu, M. *et al.* Room-Temperature Lasing in Colloidal Nanoplatelets via Mie-Resonant Bound States in the Continuum. *Nano Lett.* **20**, 6005–6011 (2020).

43. Do, T. T. H. *et al.* Room-Temperature Lasing at Flatband Bound States in the Continuum. *ACS Nano* **19**, 19287–19296 (2025).

44. Yesilkoy, F. *et al.* Ultrasensitive hyperspectral imaging and biodetection enabled by dielectric metasurfaces. *Nat. Photonics* **13**, 390–396 (2019).

45. Romano, S. *et al.* Label-free sensing of ultralow-weight molecules with all-dielectric metasurfaces supporting bound states in the continuum. *Photonics Res.* **6**, 726 (2018).

46. Doeleman, H. M., Monticone, F., Hollander, W. Den, Alù, A. & Koenderink, A. F. Experimental observation of a polarization vortex at an optical bound state in the continuum. *Nat. Photonics* **12**, 397–402 (2018).

47. Lv, J. *et al.* Optical switching with high-Q Fano resonance of all-dielectric metasurface governed by bound states in the continuum. *Opt. Express* **32**, 28334 (2024).

48. Chen, H. & Hsu, C. W. Bound States in the Continuum in Fiber Bragg Gratings. *ACS Photonics* **6**, 2996-3006 (2019).

49. Jin, R. *et al.* Toroidal Dipole BIC-Driven Highly Robust Perfect Absorption with a Graphene-Loaded Metasurface. *Nano Lett.* **23**, 9105–9113 (2023).

50. Deng, Y. *et al.* Quad-narrowband perfect absorption in near infrared for optical switching and sensing based on quasi-bound states in the continuum. *Phys. Chem. Chem. Phys.* **27**, 5843–5853 (2025).

51. Qi, K., Ge, J., Shen, X. & Gao, Y. High-Q perfect light absorption enabled by degenerate merging BICs. *Opt. Express* **33**, 29869 (2025).

52. Buchnev, O., Belosludtsev, A., Reshetnyak, V., Evans, D. R. & Fedotov, V. A. Observing and controlling a Tamm plasmon at the interface with a metasurface. *Nanophotonics* **9**, 897–903 (2020).

53. Tripathi, D. & Hegde, R. S. Phase change material metasurface loading enables an ultrafast all-optically switchable, compact, narrowband freespace optical filter. *Opt. Commun.* **569**, 130788 (2024).

54. Mohammadi Dinani, H. & Mosallaei, H. Active Tunable Pulse Shaping Using $MoS_2$-Assisted All-Dielectric Metasurface. *Adv. Photonics Res.* **4**, (2023).

# Supplementary Information

## S1. Cross-sectional SEM image after FIB ion etching

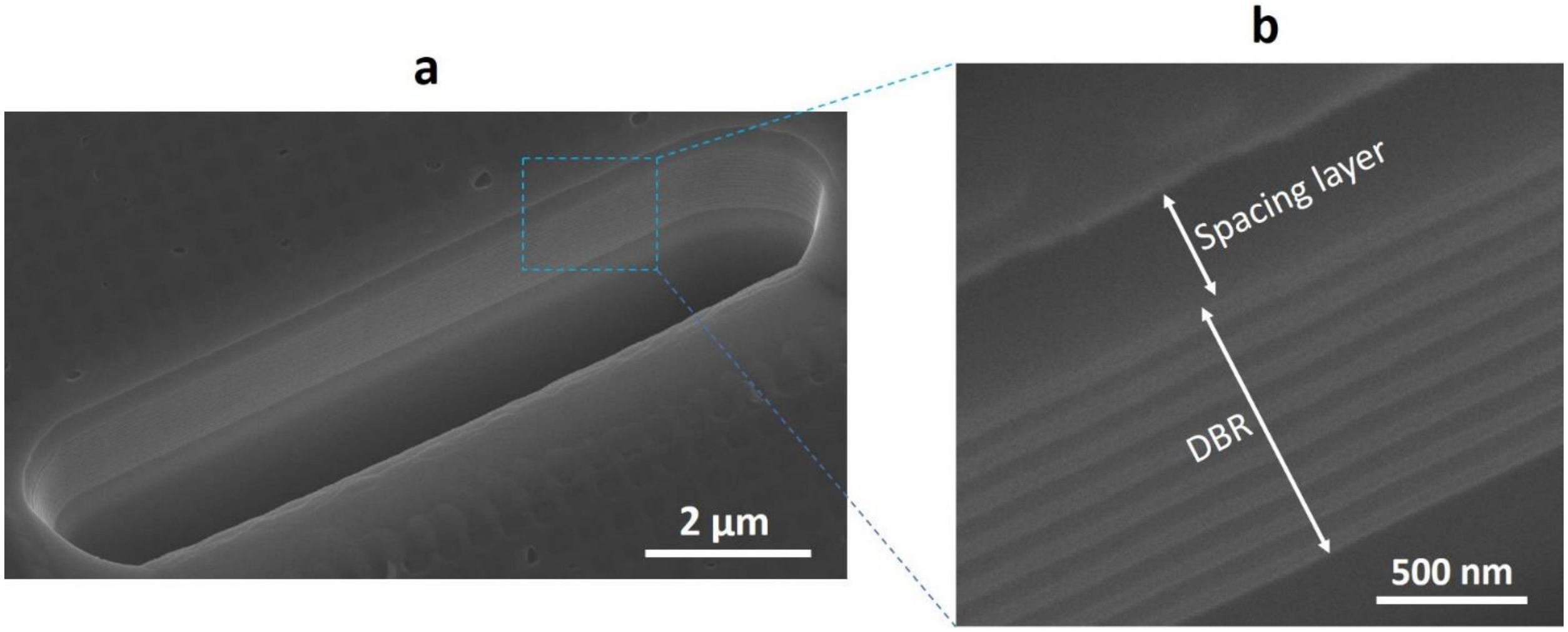

**Figure S1. Focused Ion Beam (FIB) ion etching and SEM images. a,** Cross-sectional SEM image (25° inclined with vertical direction) after gallium ion-assisted FIB etching of 10 long, 2.5 wide, and 2 μm deep cuts. The ion beam heavily damages the top metasurface layer (square-hole design). **b,** Magnified SEM image of cross-section with the DBR and the spacing layer.

## S2. Numerical reflectance without the insertion of the spacing layer

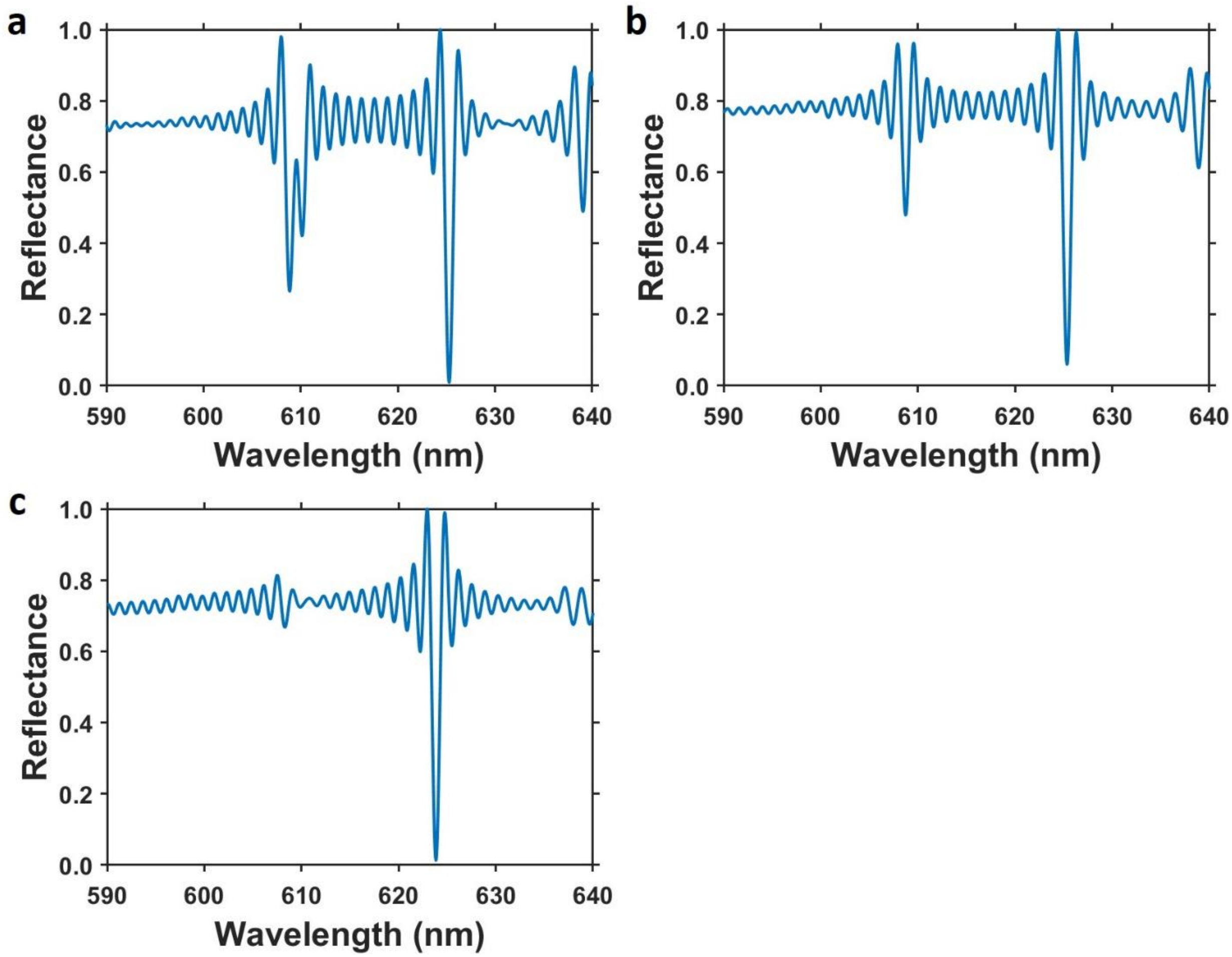

**Figure S2. Numerical reflectance spectra for the normally applied source from the device without the insertion of the $SiO_2$ spacing layer. a, b,** Reflectance spectra from the asymmetric T-shaped hole metasurface for the polarization direction of the applied source along the x-axis (a) and y-axis (b). **c,** Reflectance spectra from the square hole metasurface for both x and y-polarization.

We numerically investigated the reflection spectra of both asymmetric T-shaped and square-hole metasurfaces, without inserting a spacing layer between the DBR and the metasurfaces. Figure S2 shows that the reflectance (wavelength from 590 to 640 nm) changed significantly in the absence of the spacing layer (Spacer thickness 0 from Figure 3c). The ED resonance disappears completely as there is no coupling between the metasurface and the DBR. Besides, a high-Q absorption resonance appears at 625 nm (Figure S2a and S2b) and 624 nm (Figure S2c) without mode splitting due to the Mie resonance from the metasurface.

## S3. Cross-sectional electric field maps at the resonance dips

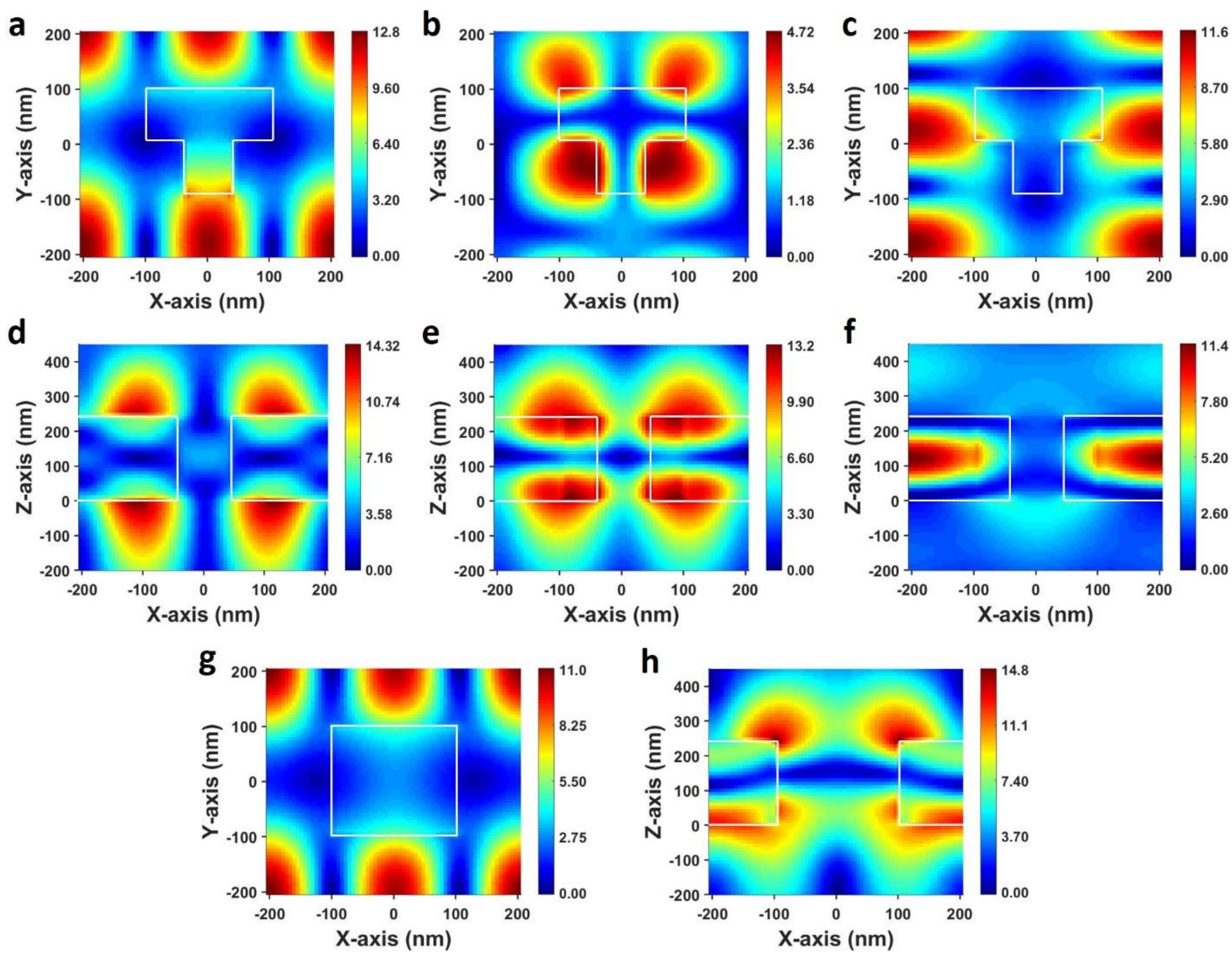

**Figure S3. Cross-sectional electric field map from the asymmetric T-shaped hole (a–f) and symmetric square hole (g, h) metasurface. a, d,** Horizontal and vertical cross-sectional electric field maps at the resonance point of 603 nm for the polarization direction of the applied source along the x-axis. **b, e,** Horizontal and vertical cross-sectional electric field maps at the resonance point of 619 nm for the polarization direction along the x-axis. **c, f,** Horizontal and vertical cross-sectional electric field maps at the resonance point of 603 nm for the polarization direction along the y-axis. **g, h,** Horizontal and vertical cross-sectional electric field maps at the resonance point of 601 nm from the square hole metasurface for both polarizations. The color bars in all figures indicate the degree of electric field localization within or on the surface of the metasurface relative to the applied source.

To gain insight into the resonant modes and the enhancement of electric-field localization within the metasurface cavity, we numerically analyze the electric-field distribution (Figure S3) from horizontal and vertical cross-sections of the metasurface. Figures S3a and S3d demonstrate that the electric field is mostly localized at the metasurface surface, with a maximum accumulation of 14 folds. The electric field distribution rotates by 90° (Figure S3c) when the polarization direction of the applied source is rotated to the y-axis. However, the electric field maintains a high degree of localization of around 11.6 folds despite rotation, with the field accumulating inside the metasurface (Figure S3f). From the electric field maps and the distribution of the electric field vector maps (Figure S4), we can conclude that the resonance at 603 nm is an electric dipole mode. We also observed the same electric field distribution at 605 nm, which supported the claim of mode splitting. Likewise, we investigated the identical electric field maps from the resonances at 619 and 622 nm, presented in Figures 4b and 4e. The field distribution shows a magnetic quadrupole (MQ) pattern (Figure S3b), with a field localization of 13.2 folds. The only resonance at 601 nm from the symmetric square hole shows identical field maps (Figure S3g) to those of the asymmetric T-shaped resonance at 603 nm, with a field localization of 14.8 folds. It is worth noting that the field localization is very high across all resonances, which is crucial for the photoluminescence and lasing applications. Additionally, the field localization at the BIC point is theoretically infinite, which is infeasible to demonstrate numerically.

## S4. Electric vector field distribution maps and multi-mode decomposition

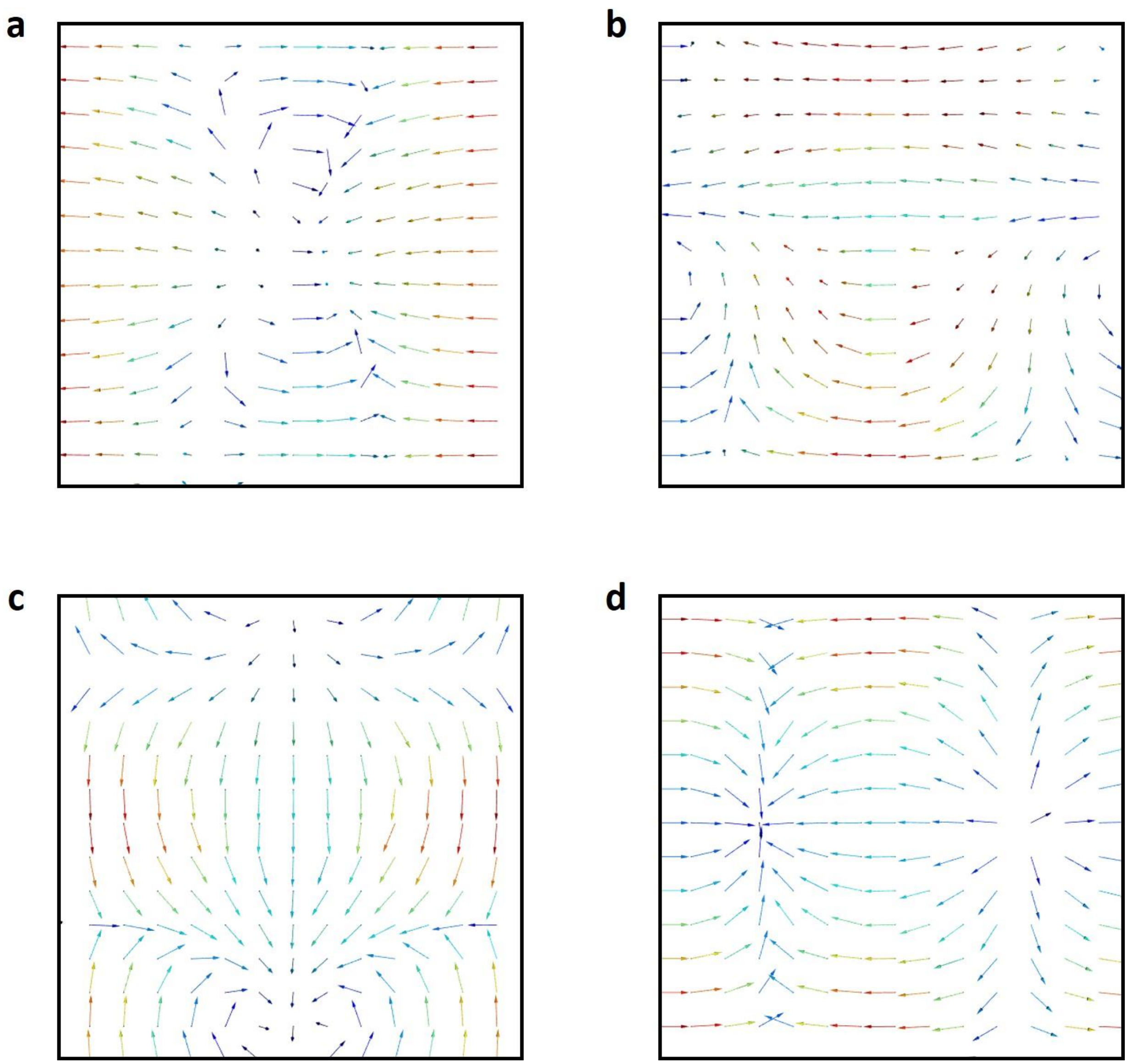

**Figure S4. Horizontal cross-sectional electric field vector distribution maps. a, b,** Vector fields from resonance point at 603 (a) and 619 nm (b) for the asymmetric T-shaped hole while the polarization direction is along the x-axis. **c,** Vector field from the resonance point at 603 nm for the asymmetric T-shaped metasurface, while the polarization direction is along the y-axis. **d,** Vector field from the resonance point at 601 nm for the square hole metasurface for both x and y-polarization directions.

We have already discussed the electric field maps and field enhancement in the previous section (S3). Here, we demonstrate the corresponding distribution of the electric field vector maps to better understand. The field distribution from Figures S4a, S4c, and S4d denotes the electric dipole (ED) mode distribution, which is aligned and rotated according to the applied electric field polarization direction. Moreover, the ED is independent of the metasurface's asymmetry. Figure S4b represents the field distribution of the magnetic quadrupole (MQ) mode, where the vector distribution and electric field map (Figure S3b) are nonuniform along the y-axis due to the asymmetry of the meta-atom. We can identify the resonances from the multipole analysis technique based on the scattering current density, J, which is defined as:

$$\boldsymbol{J} = -i\omega\varepsilon_0(\varepsilon - \varepsilon_d)\boldsymbol{E} \tag{S1}$$

where $\omega$ is the angular frequency of the applied source, $\varepsilon$ is the relative permittivity of the device material, $\varepsilon_0$ is the permittivity of vacuum, $\varepsilon_d$ is the relative permittivity of the surrounding medium of the meta-structure, and $\boldsymbol{E} = \boldsymbol{E}(\boldsymbol{r})$ is the electric field of the applied source[1]. The electric and magnetic multipoles are identified based on the Cartesian basis after decomposing the scattering current density. The electric and magnetic multipoles can be written as:

Electric dipole (ED) moment, $\boldsymbol{p} = \frac{i}{\omega}\int \boldsymbol{J}dr$ (S2)

Magnetic dipole (MD) moment, $\boldsymbol{m} = \frac{1}{2}\int[\boldsymbol{r}\times\boldsymbol{j}]dr$ (S3)

Electric quadrupole (EQ) moment, $\boldsymbol{Q_e} = \frac{i}{\omega}\int[(\boldsymbol{r}\otimes\boldsymbol{j}) + (\boldsymbol{j}\otimes\boldsymbol{r})]dr$ (S4)

Magnetic quadrupole (MQ) moment, $\boldsymbol{Q_m} = \frac{1}{3}\int[\boldsymbol{r}\otimes(\boldsymbol{r}\times\boldsymbol{j}) + (\boldsymbol{r}\times\boldsymbol{j})\otimes\boldsymbol{r}]dr$ (S5) where $\otimes$ represents the outer product[2,3].

**References**

1. Wu, M. *et al.* Room-Temperature Lasing in Colloidal Nanoplatelets via Mie-Resonant Bound States in the Continuum. *Nano Lett.* **20**, 6005–6011 (2020).

2. Kaelberer, T., Fedotov, V. A., Papasimakis, N., Tsai, D. P. & Zheludev, N. I. Toroidal dipolar response in a metamaterial. *Science.* **330**, 1510–1512 (2010).

3. Radescu, E. E. & Vaman, G. Exact calculation of the angular momentum loss, recoil force, and radiation intensity for an arbitrary source in terms of electric, magnetic, and toroid multipoles. *Phys. Rev. E.* **65**, 046609 (2002).